**This is the accepted manuscript (postprint) of the following article:**
A. Bolandparvaz Jahromi, E. Salahinejad, *Competition of carrier bioresorption and drug release kinetics of vancomycin-loaded silicate macroporous microspheres to determine cell biocompatibility*, Ceramics International, 46 (2020) 26156-26159.
https://doi.org/10.1016/j.ceramint.2020.07.112

# Competition of carrier bioresorption and drug release kinetics of vancomycin-loaded silicate macroporous microspheres to determine cell biocompatibility

A. Bolandparvaz Jahromi, E. Salahinejad *

Faculty of Materials Science and Engineering, K. N. Toosi University of Technology, Tehran, Iran

## Abstract

Bioceramic porous microspheres are promising substances for dental and orthopedic bone void filling, tissue engineering and drug delivery applications. In this research, the structure and cytocompatibility of bioactive magnesium-calcium silicate macroporous microspheres loaded with vancomycin hydrochloride, an antibiotic drug, were studied. In this regard, bredigite ($Ca_7MgSi_4O_{16}$), akermanite ($Ca_2MgSi_2O_7$) and diopside ($CaMgSi_2O_6$) carriers were fabricated through a sequence of sol-gel, calcination, droplet extrusion and sintering processes, followed by impregnated with vancomycin. X-ray diffraction (XRD) and scanning electron microscopy verified the formation of the desired ceramic crystalline phases and macroporous characteristics of the carriers, respectively. Based on the MTT assay, the antibiotic-loaded bredigite, akermanite and diopside devices comparatively exhibited the lowest, intermediate and highest levels of human bone marrow mesenchymal stem cells (hBM-MSCs) viability and proliferation, respectively. It was concluded that the role of the carrier bioresorption kinetics prevails over the drug delivery kinetics in determining the cell biocompatibility of the devices.

* Email Address: <salahinejad@kntu.ac.ir>

## 1. Introduction

Osseous defects can be treated by various surgical approaches, including the use of dense blocks, tissue-engineering scaffolds, bone grafts, microspheres and so on, depending on their size and location in the body. For small defects, microspheres have several advantages in comparison to irregular-shaped grafting powders. Typically, the spherical shape of bone void fillers provides better packing and flow properties during injection, more suitable cell attachment and adhesion and less inflammatory responses [1, 2]. A number of apatite-based microspheres have been also commercialized by, for example, Curasan, Biomatlante and Biomet companies for dental and orthopedic applications. Medical microspheres can be also equipped with nanometric and micron-sized interconnected pores, offering promising features for tissue reconstruction and the delivery of drugs and biological agents [3-5].

The risk of osteomyelitis, the infection and inflammation of bone, is always of concerns in the invasive therapies of bone defects, which can be overcome by using antibiotic delivery systems. To do so, antibiotic-loaded poly(methyl methacrylate)-based bone cements were developed, with the major drawback of non-biodegradability [6-8]. Accordingly, the employment of bioresorbable and bioactive ceramics in antibiotic delivery systems started to grow for bone defect treatment. In this regard, magnesium-calcium biosilicates are under special consideration due to their considerable osteoinductivity, osteoconductivity and bioresorbability [9, 10].

Regarding drug-loaded magnesium-calcium silicate microspheres, diopside microspheres impregnated with dexamethasone exhibited the attractive potential for bone tissue regeneration [11]. Furthermore, the release kinetics of vancomycin, an orthopedic antibiotic drug, from Mg–Ca silicate microspheres was recently reported to be controlled by combinations of diffusion and degradation mechanisms [12]. The aim of the current work is to compare the cell cytocompatibility of bredigite ($Ca_7MgSi_4O_{16}$), akermanite ($Ca_2MgSi_2O_7$) and diopside ($CaMgSi_2O_6$) macroporous microsphere bone void fillers loaded with vancomycin. In this regard, the contribution of the carrier bioresorption and drug release kinetics to the cell viability is distinguish.

## 2. Materials and methods

The sol-gel technique was utilized to synthesize Ca-Mg silicate powders, using calcium nitrate tetrahydrate ($Ca(NO_3)_2.4H_2O$, Merck, >98%), magnesium nitrate hexahydrate ($Mg(NO_3)_2.6H_2O$, Merck, >98%) and tetraethyl orthosilicate ($(C_2H_5O)_4Si$, TEOS, Merck, >98%) precursors. First, TEOS was loaded into an aqueous solution of nitric acid ($HNO_3$, Merck, 65%) with the molar ratio of TEOS/ $HNO_3$/ $H_2O$ = 1: 0.16:8. After 30 min of stirring for hydrolysis, in accordance with the bredigite ($Ca_7MgSi_4O_{16}$), akermanite ($Ca_2MgSi_2O_7$) and diopside ($CaMgSi_2O_6$) molar stoichiometries, the proper amounts of $Ca(NO_3)_2.4H_2O$ and $Mg(NO_3)_2.6H_2O$ were added and stirred for 5 h. The obtained solutions were then aged at 60 °C for 24 h, dried at 120 °C for 48 h and finally calcined at 1300 °C for 3 h. X-ray diffraction (XRD, PHILIPS-PW1730) was used for the phase analysis of the calcined powders.

The prepared silicate powders were used as feedstocks to produce macroporous microspheres by a droplet extrusion process using carbon porogens. For this purpose, the silicate and spherical carbon (Sigma Aldrich, >99.95%) powders with the mass ratio of 2:3 were loaded to a 3 wt% sodium alginate ($NaC_6H_7O_6$, Sigma Aldrich) aqueous solution. After sonication for 10 min, the suspensions were extruded via a syringe into an aqueous solution of calcium chloride ($CaCl_2$, Merck, >98%) for alginate cross-linking. The precipitated microspheres were afterwards washed, dried, and eventually sintered at 1000 °C for 2 h. A scanning electron microscope (SEM, TESCAN, MIRA3, accelerating voltage = 15 kV) was used to observe inner pores created as a result of the carbon porogen burning.

The fabricated microspheres were loaded with vancomycin by exposing to a 0.05 % solution of vancomycin hydrochloride ($C_{66}H_{75}Cl_2N_9O_{24}$, Sigma Aldrich) at the microsphere/drug mass ratio of 120:1, followed by drying at 60 °C.

The cytocompatibility of the antibiotic-loaded microspheres with respect to human bone marrow mesenchymal stem cells was evaluated by the MTT assay. The samples were sterilized in a 70 % solution of ethanol, rinsed with the phosphate buffered saline solution and then exposed to ultraviolet radiation. 5000 cells were seeded onto 600 µg of the sterilized microspheres in Dulbecco's modified Eagle's medium glucose supplemented with fetal bovine serum and pen/strep. The control sample was the cell-containing medium without any microspheres. After incubation for 1, 3 and 7 days, the MTT assay was conducted along with measuring the optical density of viable cells by a Elisa microplate reader at 630 nm. The variance analysis with a significance level (p-value) less than 0.05 was considered to compare the data with three repetitions.

## 3. Results and discussion

The XRD patterns of the sol-gel derived powders after calcination at 1300 °C are presented in Figure 1. According to the XRD peak analysis by the PANalytical X'Pert HighScore software, stoichiometric diopside (PDF Ref. code: 017-0318), akermanite (PDF Ref. code: 035-0592) and bredigite (PDF Ref. code: 036-0399) monophase structures have been desirably developed in the related samples, suggesting the merit of the sol-gel and calcination processes utilized. Using the Scherrer equation [13], the mean crystallite size of the bredigite, akermanite and diopside phases is estimated to be almost 57, 35 and 48 nm, respectively. The melting point of bredigite, akermanite and diopside and is 1372, 1454 and 1391°C, respectively [9]. Accordingly, the different crystallite sizes of the calcined phases are attributed to their melting points, where the higher homologous temperature the faster crystallite growth at the calcination temperature used for all the samples.

Figure 2 shows the SEM micrograph of the inner section of a randomly selected microsphere sintered at 1000 °C. It is characterized by uniformly-distributed interconnected pores of 80-2000 nm is size. The pore interconnectivity of the carriers provides an efficient pathway for nutrient transport and waste removal [1, 3] and a considerable surface area for drug adsorption [2, 14], which is essential for bone tissue regeneration and drug delivery applications, respectively. The regular microsphere geometry of these porous silicate carriers fabricated by the same sequence of the sol-gel, calcination, droplet extrusion and sintering processes has been previously ascertained [12].

Figure 3 indicates the results of the MTT assay on human bone marrow mesenchymal stem cells cultured on the vancomycin-loaded microspheres. Considering $p$-value $< 0.05$, there are no statistically significant differences between the control and prepared devices for the first day of culture. However, for the more prolonged durations of the cell culture, the cytocompatibility of the drug-loaded samples ranks as diopside > akermanite > bredigite, where the bredigite, akermanite and diopside systems exhibit the lower, equivalent and higher cell viability than the control, respectively. Given the similar geometry of the fabricated carriers, the different cell biocompatibility levels of the microspheres can be explained in terms of the kinetics of the drug delivery and ions release from the bioresorbable microspheres, as follows.

On the one hand, the mechanism of vancomycin delivery from these microspheres has been reported to be diffusion-controlled at short periods [12], reflecting the negligible bioresorbability of the carriers at these periods. In addition, almost 85, 72 and 60 % of the drug loaded amounts are released from the bredigite, akermanite and diopside devices, respectively, at the first 24 h of exposure [12]. Ignoring the role of the carries resorption and considering the relatively high and unequal levels of the drug released from the different samples, the cell viability equivalency of the three microspheres denies the deteriorous role of the drug release in cytocompatibility for this period of time. This is attributed to the safe type and level of the drug loaded despite these burst releases from the viewpoint of cell biocompatibility, where the burst release of some drugs from some drug delivery devices has been reported to deteriorate biocompatibility due to the drug shock [15]. It should be albeit noted that the initial burst release of antibiotics is required to treat osteomyelitis because it can inhibit the initial adherence of bacterial and the subsequent biofilm formation [16, 17].

On the other hand, the drug release increments of less than 10 % at a sustained mode have been detected from the first day to the seventh day of immersion for the same drug delivery devices [12]. Since the burst releases of 85, 72 and 60 % during one day did not impose any negative effect on cytocompatibility, the sustained release increment of as small as 10 % during six days is postulated to be not detrimental. It is accordingly inferred that the bioresorption of the carriers and thereby ions release, rather than vancomycin release, mainly determine the cell biocompatibility difference of the samples at the third and seventh days of the cell culture. This is supported by the fact that the contribution of the bioceramics bioresorption prevails over diffusion to the drug delivery kinetics at the longer periods of soaking [12]. The non-detrimental effect of vancomycin on cytocompatibility realized in this work is also compatible with Refs. [18, 19]. It would be also worth mentioning that the cytocompatibility ranking of the drug delivery systems developed in this work is in good agreement with the comparison reported on the osteoblast cytocompatibility of bredigite, akermanite and diopside drug-unloaded disks [20]. Typically, by decreasing the silicon content in the order diopside, akermanite and then bredigite, the structural stability against bioresorption is reduced. This subsequently leads to the excess release of ions, the shift of physiological pH to alkalosis, and deterioration of biocompatibility.

## 4. Conclusions

In this work, bredigite, akermanite and diopside macroporous microspheres were successfully fabricated by the sol-gel and droplet extrusion processes for bone void filling applications. Afterwards, they were impregnated with vancomycin via immersion in a solution of the drug and studied in terms of in vitro cytocompatibility with respect to human bone marrow mesenchymal stem cells. The results showed that the cell viability on the drug-loaded devices ranks as diopside > akermanite > bredigite, so that these systems exhibited the higher, equivalent and lower biocompatibility than the control, respectively. Considering the drug delivery kinetics from the carriers, it was also concluded that the contribution of the carriers bioresorption to the cell cytocompatibility prevails over the drug release kinetics.

## Figures

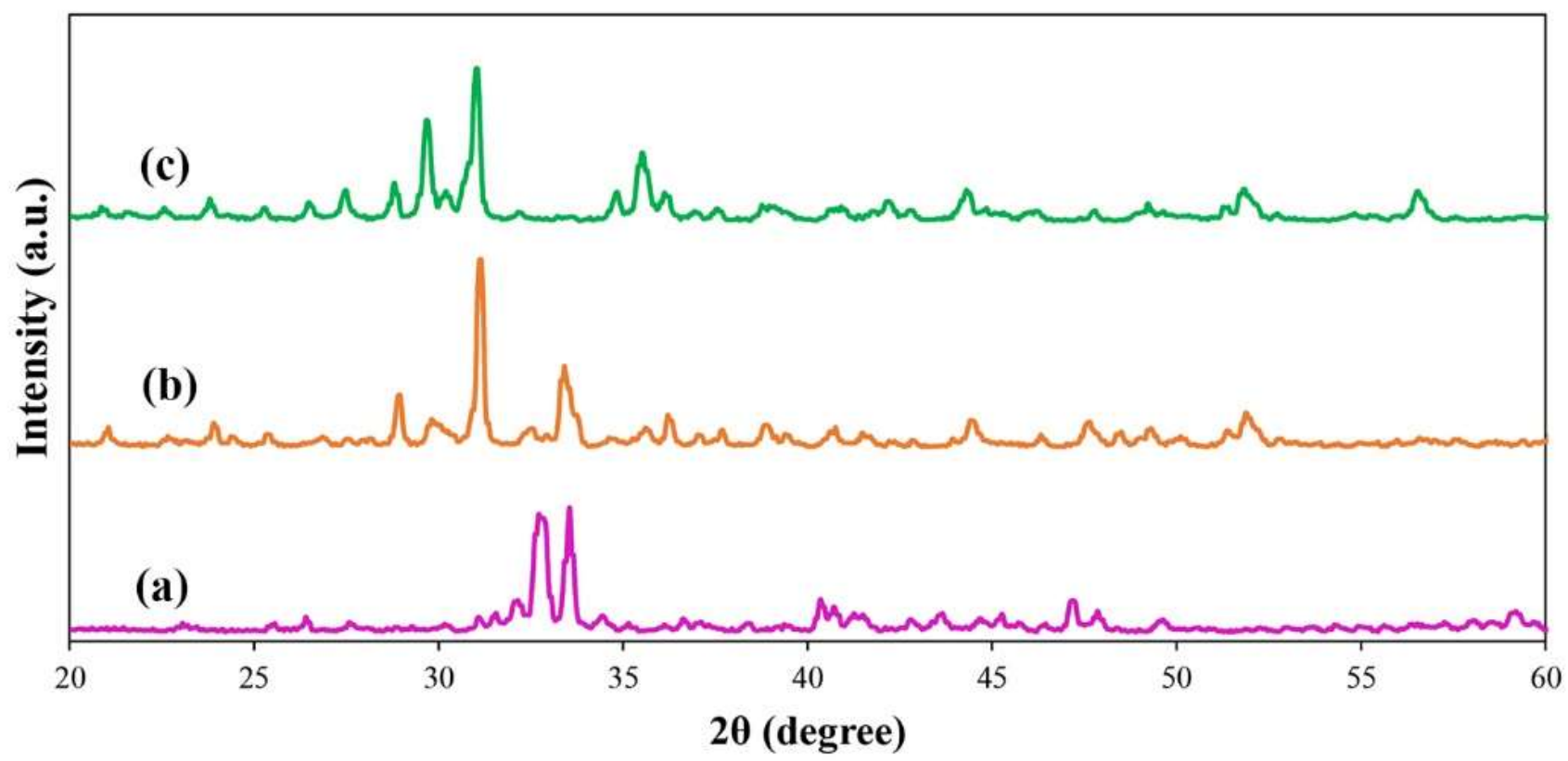

Figure 1. XRD patterns of the powder samples loaded in the bredigite (a), akermanite (b) and diopside (c) stoichiometries after calcination at 1300 °C, in which all peaks are related to the bredigite ($Ca_7MgSi_4O_{16}$), akermanite ($Ca_2MgSi_2O_7$) and diopside ($CaMgSi_2O_6$) phases, respectively.

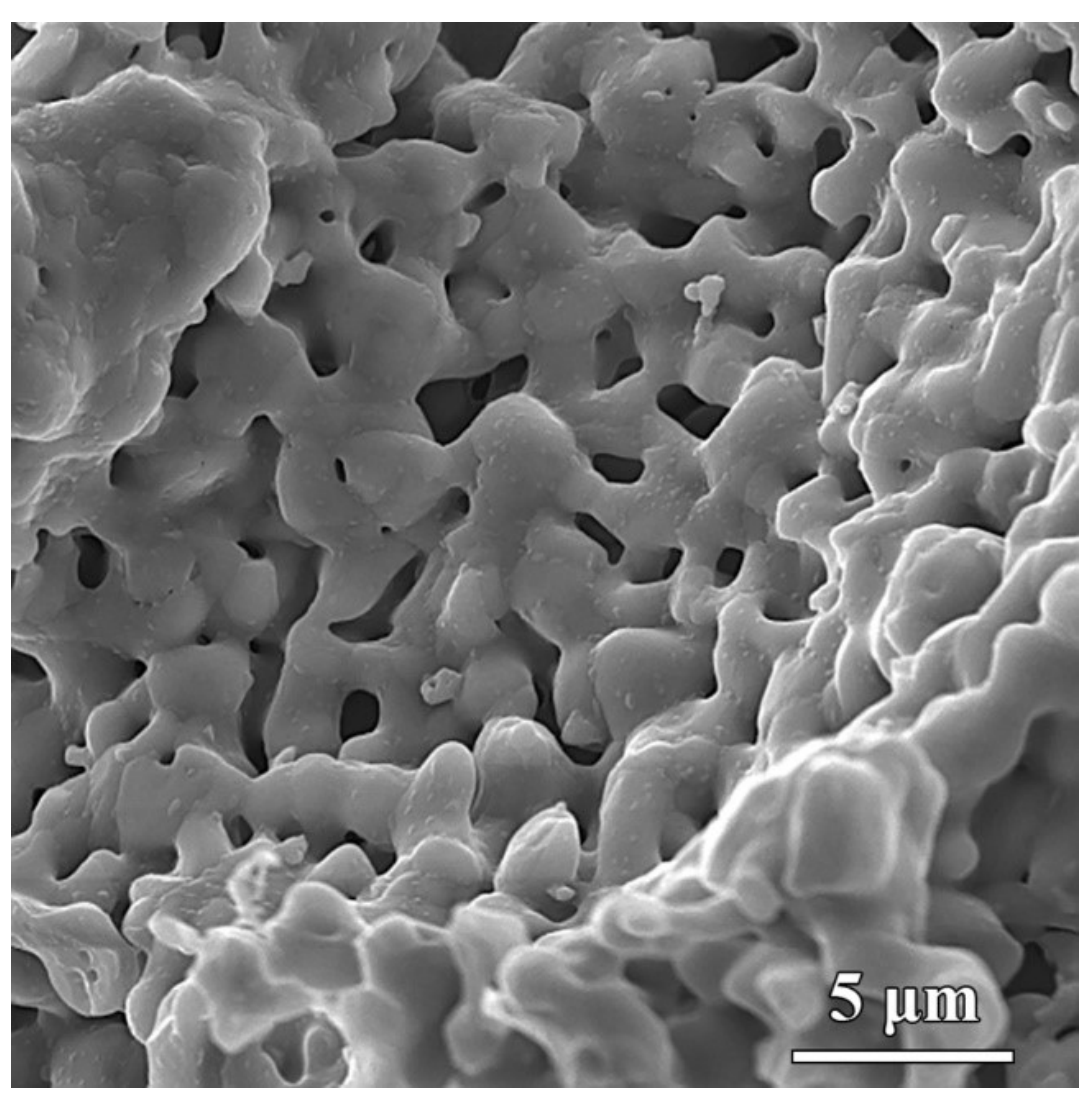

Figure 2. SEM micrograph of the inner structure of a microsphere sintered at 1000 °C.

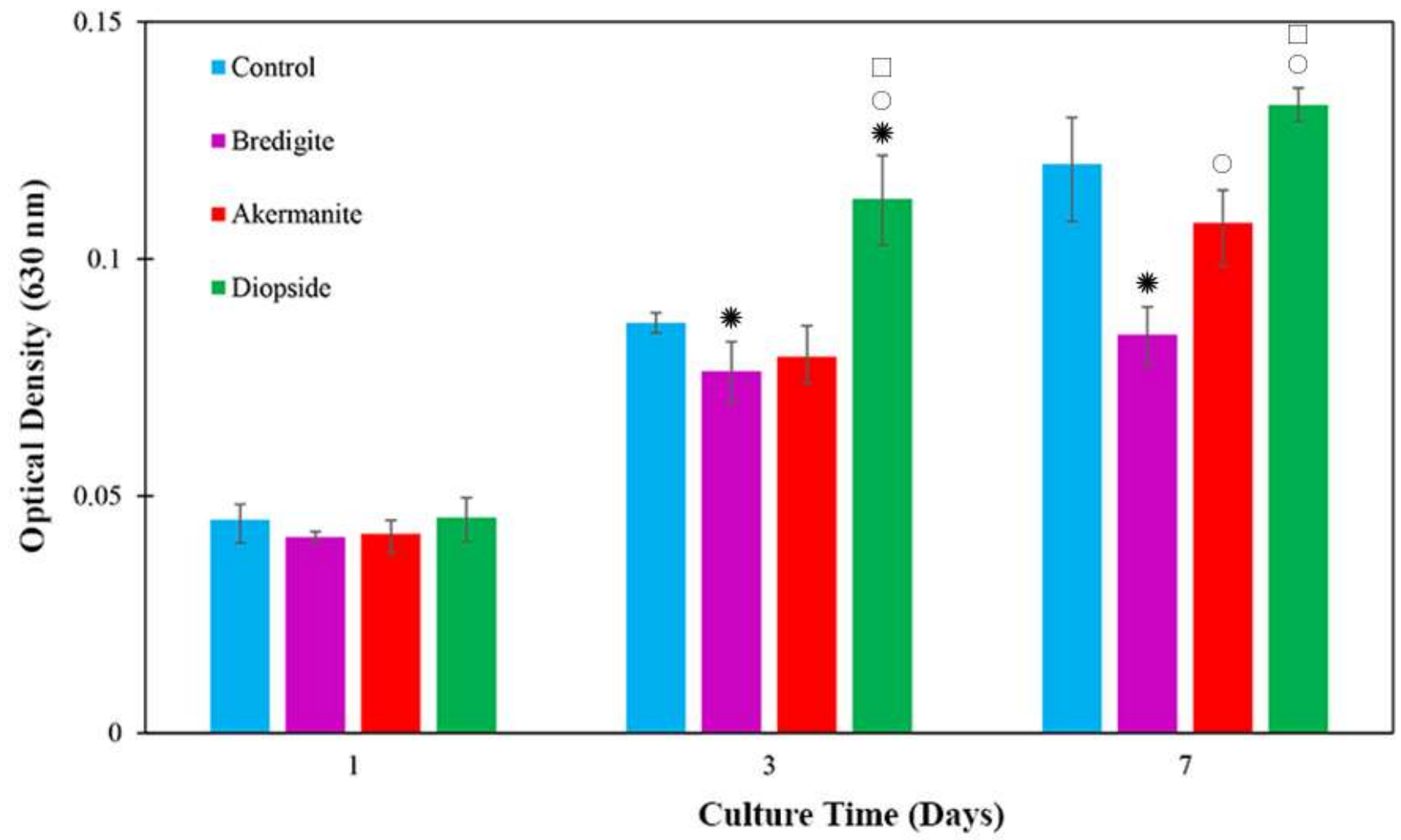

Figure 3. MTT assay results of the cell culture on, from left to right, the control, bredigite, akermanite and diopside drug-loaded microspheres. *, ○ and □ represent significant differences with respect to the control, bredigite and akermanite vancomycin-impregnated samples, respectively ($p < 0.05$).